\documentclass{emulateapj}
\usepackage{graphics,epsfig}
\usepackage{graphicx}
\usepackage{amsmath}
\usepackage{natbib}
\usepackage{amssymb}
\usepackage{longtable}
\usepackage{times}
\slugcomment{\copyright 2007. The American Astronomical Society. All rights reserved.}

\shorttitle{Turbulence scale size in comet plasma tail}
\shortauthors{Roy, Manoharan, \& Chakraborty}

\begin{document}
\title{Occultation observation to probe the turbulence scale size in the plasma 
tail of comet Schwassmann-Wachmann 3-B}

\author{Nirupam Roy\altaffilmark{1}, P. K. Manoharan\altaffilmark{2}, and 
Pavan Chakraborty\altaffilmark{3}}

\altaffiltext{1} {NCRA-TIFR, Post Bag 3, Ganeshkhind, Pune 411 007, India; 
nirupam@ncra.tifr.res.in}
\altaffiltext{2} {Radio Astronomy Centre, NCRA-TIFR, PO Box 8, Udhagamandalam 
643 001, India; mano@ncra.tifr.res.in}
\altaffiltext{3} {Indian Institute of Information Technology, Deoghat, Jhalwa, 
Allahabad 211 012, India; pavan@iiita.ac.in}

\begin{abstract}
{We present the occultation observation of compact radio source B0019$-$000 
through the plasma tail of comet Schwassmann-Wachmann 3-B. The observation 
was made with the Ooty Radio Telescope at 326.5 MHz on 2006 May 26, when the 
plasma tail of the comet was in front of this source. The scintillation was 
found to be increased significantly for the target source compared to that 
of a control source. The intensity fluctuation power spectra show both 
steepening at high spatial scales and excess power at low spatial scales. 
This observation can be attributed to the turbulence in the comet plasma tail. 
A two-regime plasma turbulence can explain the time-evolution of the power 
spectrum during the occultation observation.}
\end{abstract}


\section{Introduction}
\label{sec1}

The interplanetary scintillation (IPS) technique is a useful, convenient, 
remote-sensing method to probe the solar wind \citep{hw64,mano93}. It exploits 
the scattering of radio waves from a compact radio source by the density 
irregularities in the solar wind and provides information on the speed and 
turbulence spectrum of the solar wind plasma at the region of closest solar 
offset of the line of sight to the radio source \citep[e.g.,][]{mano93}. When 
a comet plasma tail intersects the radio path, the electron density 
irregularities associated with the comet can cause an enhancement in the 
level of scintillation. Several attempts have been made to estimate 
the turbulence associated with the cometary plasma by observing the occultation 
of radio source by the tail of comet \citep{an75,an87,al86,al89,sl87,hd87}. 
However, in order to confirm the effect of the cometary plasma as well as 
to assess the contribution to the level of turbulence imposed by any solar 
wind transients or disturbances, it is essential to simultaneously monitor 
a region just outside the tail of the comet during such observations. 

\citep{an87} reported a significant increase in the plasma turbulence as 
the tail of comet Halley approached a radio source. But, a rather similar 
enhancement in scintillation was also observed towards control sources, 
which were located outside the comet tail \citep{an87}. This emphasized 
the need for near-simultaneous monitoring of IPS through the tail of the 
comet as well as a region outside the comet tail. This study attributed 
an insignificant increase of turbulence associated with the cometary plasma. 
In a later work, \citet{sl90} adopted the idea of near-simultaneous 
observation of control source during cometary occultation observation and 
reported a significant increase of the scintillation of the target source 
compared to that of the control source. In this Letter we report the 
occultation of the compact radio source B0019$-$000 by the plasma tail of 
comet Schwassmann-Wachmann 3-B. The results are substantiate by the 
observations of the IPS of a control source outside the tail. Systematic 
changes observed in the IPS spectra as the source approached the comet 
tail suggest a steeper turbulence spectrum for the cometary plasma at scales 
$>$500 km and excessive turbulence at the small-scale part (i.e., $<$50 km) of 
the spectrum. Below, in \S2, we give the details of our observations and the 
results. We compare the results with previously reported results and discuss 
the actual astrophysical implication of the results in \S3. Finally, we 
summarize and present our conclusions in \S4. 

\section{Observations and results}
\label{sec2}

The comet Schwassmann-Wachmann 3 (also known as 73P/Schwassmann-Wachmann) 
is a periodic comet. This Jupiter-family comet passed within $\sim$0.1 AU of 
the Earth around the middle of 2006. It is to be noted that the comet was going 
through breaking and disintegration \citep{bk95,cro96,bo02}, which would 
likely to increase the level of turbulence. The occultation observation 
reported here was therefore aimed at estimating the turbulence spectrum of 
density irregularities and associated scale sizes of the cometary plasma. 

Compact sources lying close to the path of the comet were chosen from the 
Molonglo Catalogue \citep{lrg81}.  On 2006 May 26, one of the fragments 
(labeled 73P-B, the second brightest fragment at the time of observation) 
aligned along the line of sight to the radio source B0019$-$000, which has a 
compact component ($\sim$ 50 mas) containing $>75\%$ of the total flux 
density, S$_{\rm 408}$=2.99$\pm$0.14 Jy. The line of sight along the radio 
source B0019$-$073 (S$_{\rm 408}$ = 1.83$\pm$0.07 Jy), was monitored as the 
control region. The systematic monitoring of these sources on the day of 
occultation as well as for the next two consecutive days provide the spectrum 
of the density irregularities and dominant scale sizes of the cometary plasma. 
This Target of Opportunity observation was carried out using the Ooty Radio 
Telescope (ORT) at the Radio Astronomy Centre, Ooty \citep{gs71}. The ORT, an 
equatorially mounted parabolic cylindrical antenna, is 530 m long in the 
north-south direction and 30 m wide in east-west direction. It operates at 
326.5 MHz and can effectively probe solar wind in the heliocentric distance 
range of 0.05 -- 1.0 AU. 

\begin{figure}
\epsscale{0.95}
\includegraphics[scale=0.65, angle=0.0]{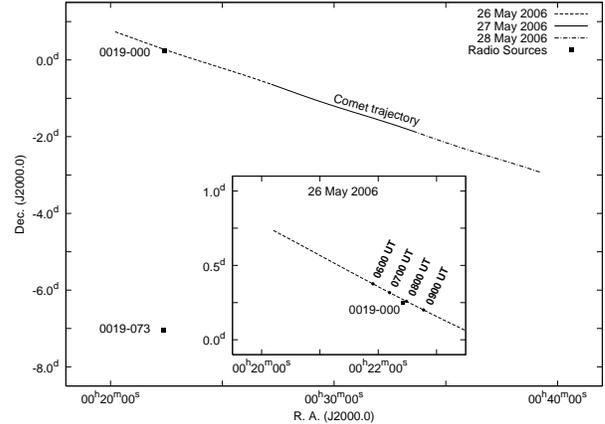}
\caption{\label{f1} \small{Path of the comet with respect to the radio 
sources (both the target and the control source). The position of the comet 
nucleus on2006 May 26 -- 28 is plotted. Part of the trajectory on 2006 May 26 
is shown in the inset, where the positions close to the time of 
observation are marked.}}
\end{figure}

\begin{figure}
\epsscale{0.95}
\includegraphics[scale=0.45, angle=-90.0]{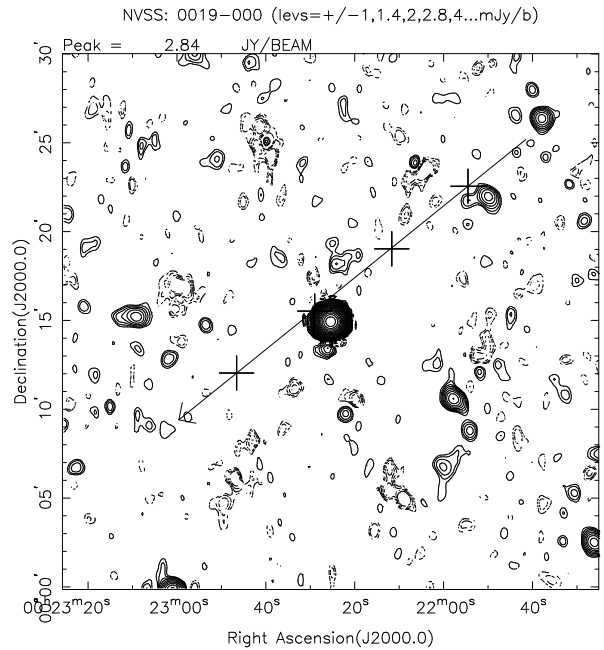}
\caption{\label{f2} \small{NVSS L-band image of the field centered at the 
target source 0019$-$000. Part of the trajectory on 2006 May 26 is 
overplotted with the position at 0600, 0700, 0800, and 0900 UT (top right 
to bottom left) marked with crosses.}}
\end{figure}

On 2006 May 26 -- 28, both the target source (B0019$-$000) and the control 
source (B0019$-$073) were observed. The coordinates and flux densities of 
these sources are given in Table (\ref{tab:tab1}). The orbital parameters of 
the comet Schwassmann-Wachmann 3-B are summarized in Table (\ref{tab:tab2}). 
Fig. (\ref{f1}) shows the position of the nucleus of the comet on 2006 May 26 
-- 28 with respect to the position of both the target and control sources, 
projected on the plane of the sky.  Fig. (\ref{f2}) displays the comet 
trajectory on 2006 May 26 over-plotted on the 1.4 GHz image of the field 
centered at the target source B0019$-$000 taken from the NRAO VLA Sky Survey 
\citep[NVSS;][]{co98}. The positions of the comet on these days are taken from 
the online high-precision ephemerides provided by Jet Propulsion Laboratory 
(JPL) solar system dynamics group. The target source was observed during 
occultation on 2006 May 26 from 0713 to 0804 UT (when the target hour angle 
reached the west limit of the ORT) at 326.5 MHz with a total bandwidth of 4.0 
MHz and a time resolution of 20 ms. During this span, the control source was 
observed for a short period to understand the solar wind characteristics and 
to distinguish between any changes caused by the turbulence in the tail of the 
comet and the solar wind. Note that the control source is sufficiently close 
to the target source. So, these two lines of sight are expected to have 
similar solar wind characteristics. A nearby cold region of the sky was also 
observed to get the ``off-source'' reference. Both the target and the control 
sources were observed again on 2006 May 27 and 28. 

\begin{table}
  \caption{Radio sources parameters}
  \begin{tabular}{cccc}
  \hline
\multicolumn{2}{c}{IAU Name}& Ecliptic coordinates & S$_{\rm 408}^a$\\
       B1950   &   J2000    & J2000 ($\deg$)       &     (Jy)       \\
  \hline          
    0019$-$000 & 0022$+$002 & 5.245 $-$ 1.998      &  2.99$\pm$0.14 \\
    0019$-$073 & 0022$-$070 & 2.320 $-$ 8.685      &  1.83$\pm$0.07 \\
  \hline
\multicolumn{4}{l}{$^a$ From Molonglo Reference Catalogue \citep{lrg81}}
\label{tab:tab1}
\end{tabular}
\end{table}

\begin{table}
  \caption{Comet 73P/Schwassmann-Wachmann 3-B parameters$^a$}
  \begin{tabular}{cccccc}
  \hline
Date    & \multicolumn{2}{c}{Nuclear ecliptic} & Heliocentric & Topocentric & Elongation \\
at      & \multicolumn{2}{c}{ coordinates }    & Distance  & Distance & Leading \\
0800 UT &  ($\deg$) &  ($\deg$) &  (AU) & (AU) & ($\deg$) \\
  \hline
May 26  &  251.15 &  $-$0.25  & 0.957 & 0.1216 & 59.62 \\
May 27  &  252.48 &  $-$0.52  & 0.955 & 0.1287 & 59.60 \\
May 28  &  253.82 &  $-$0.79  & 0.952 & 0.1360 & 59.70 \\
  \hline
\multicolumn{4}{l}{$^a$ From JPL solar system dynamics group}
\label{tab:tab2}
\end{tabular}
\end{table}

A detailed account of the IPS observation using the ORT and data analysis 
procedure is given in \citet{mano03}. In brief, the Fourier transformation 
and spectrum computation from the recorded intensity scintillation 
measurements are performed to get the power spectrum of the intensity 
fluctuations to bring out the statistical properties of the intervening 
medium. A 2048-point fast Fourier transformation yields a frequency resolution 
of $\approx$0.025 Hz. The spectrum is corrected for the off-source noise 
and the averaging of the adjacent spectral points gives a uniform noise up 
to the cutoff frequency (where the power drops to noise level). The area 
under the power spectrum is the measure of the scintillation index. Since the 
present measurements are made in the weak scattering region, the shape of the 
spectrum and the scintillation index are directly related to the turbulence 
spectrum of the intervening plasma and the level of plasma turbulence, 
respectively \citep{mano93}. 

\begin{figure*}
\epsscale{0.95}
\includegraphics[scale=0.65, angle=-90.0]{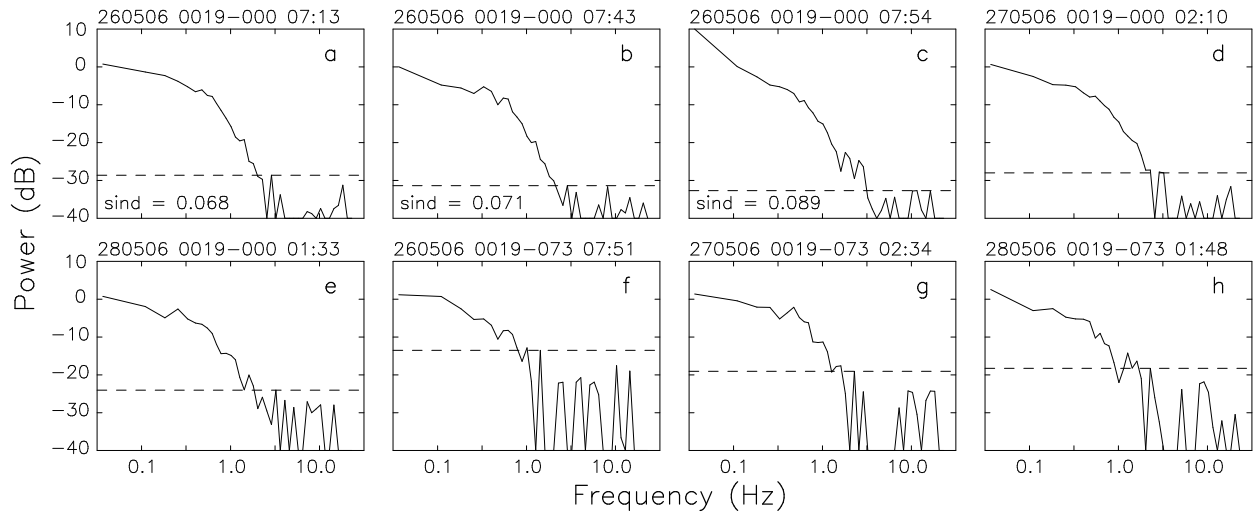}
\caption{\label{f3} \small{Power spectra of the intensity fluctuation. The 
horizontal axis is the spectral frequency (Hz) and the vertical axis is power 
(dB). Date, source name, and the time of observation (UT) are given at the top 
of the each panels. The dashed lines indicate the noise power level. Note the 
steepening of the spectrum at high spatial scales (panel $b$ and $c$) and 
significant excess power at low spatial scales (panel $c$) compared to the rest 
of the spectra from follow-up and control observations (See \S{\ref{sec2}} for 
details).}}
\end{figure*}

\begin{figure}
\epsscale{0.95}
\includegraphics[scale=0.45, angle=0.0]{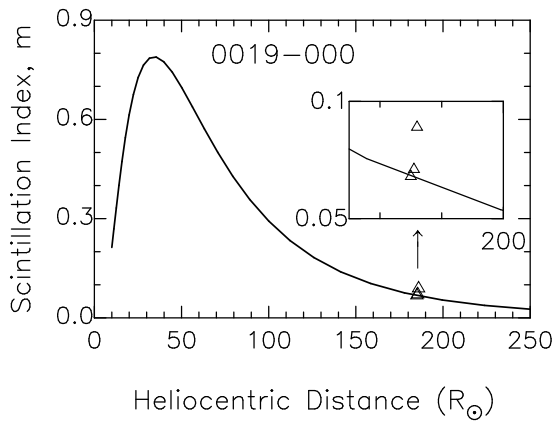}
\caption{\label{f4} \small{Scintillation index enhancement in occultation. The 
solid line represents the long-time average of the scintillation index as a 
function of heliocentric distance of the point of closest approach of the line 
of sight towards 0019$-$000 obtained from scintillation monitoring for many 
years from the ORT. The triangles are measurements during occultation. The 
part of the plot of our interest is magnified in inset.}}
\end{figure}

As illustrated in Fig. (\ref{f3}), significant changes are evident in the 
shape of the power spectrum as the comet nucleus approached the line of sight 
to the radio source. The time of observation (in UT) for each spectrum is 
given on the top right of each panel. The details of the implication of the 
power spectrum shape and the interpretation of scale size from IPS spectral 
variation are described in \citet{mano94} and \citet{mano00}. Panels $a$ -- 
$c$ in Fig. (\ref{f3}) show the change of the power spectra during occultation 
on 2006 May 26. Panels $d$ and $e$ are the power spectra from the follow-up 
observations on 2006 May 27 and 28, respectively. The power spectra for the 
control source taken on May 26 (near simultaneous to occultation observation), 
27, and 28 are shown in panels $f$ -- $h$, respectively. The spectrum 
towards the source obtained from the first half of the observation (0713 -- 
0740 UT) on May 26 has no significant change. But spectra from the later 
part of the observation (0743 -- 0804 UT) shows steepening at lower spatial 
frequency. The scale size $a=v/(2 \pi f)$, where $f$ is the temporal 
frequency and $v$ is the solar wind speed. Considering a typical solar wind 
speed of $\sim$400 km s$^{-1}$, this translates to spatial scales of $>500$ 
km. Excess power is also seen at low spatial scales ($<50$ km) on the spectrum 
obtained from the end part of the observation (0754 -- 0804 UT). The shape of 
the power spectra of the control observations shows normal quiet solar wind 
condition. The changes in the spectra during the occultation are attributed to 
the turbulence in the comet plasma tail. 

We find that the scintillation index of the control source was normal 
during these observations and also during the control observations on the 
following days, implying normal quiet solar wind conditions. But we have 
noticed significant changes in the scintillation index for the target source. 
The scintillation index increased, in about 1 hr time during the course of 
our observation, from 0.068 to 0.089. It is clear from the power spectra that 
the steepening at the high spatial scales is the dominant source of this 
enhancement. Here we consider the ratio of the level of the electron density 
spectrum $\Phi_{\rm ne}$ to the mean squared electron density $\langle 
n_{\rm e}^2\rangle$ to be a measure of plasma turbulence. For an intervening 
medium of effective thickness $L$, the scintillating flux density variance 
$\Delta S^2 \propto \Phi_{\rm ne}L$. From this, we can relate the change of 
measured scintillation index to the physical parameters of solar wind and 
cometary plasma, 
\begin{equation}
\label{eqn:eqn1}
\frac{\Delta S^2}{\langle\Delta S^2\rangle}=\frac{(\Phi_{\rm ne}/n_{\rm 
e}^2)_{s} n^2_{\rm es} L_{\rm s} + (\Phi_{\rm ne}/n_{\rm e}^2)_{\rm c} 
n^2_{\rm ec} L_{\rm c}}{(\Phi_{\rm ne}/n_{\rm e}^2)_{\rm s} n^2_{\rm es} 
L_{\rm s}} 
\end{equation}
where the subscripts `s' and `c' refer to solar and cometary, respectively, and 
$\langle\Delta S^2\rangle$ is the estimated variance due to the solar wind on 
that day.  

\begin{table}
  \caption{Parameters for the plasma tail on May 26}
  \begin{tabular}{ccccccc}
  \hline
 Time  & Flux$^a$ & Scintillation &  $\Delta S^b$ & $\Delta S^2$ & & $(\Phi_{\rm 
ne}/n_{\rm e}^2)_{\rm c}$\\
\cline{5-5} \cline{7-7}
  (UT) &  (Jy)    &  index        &     (mJy)   & $\langle\Delta 
S^2\rangle$ & & $(\Phi_{\rm ne}/n_{\rm e}^2)_{\rm s}$\\
  \hline          
0725 & 3.1 & 0.068 & 210.8 & 1.327 & &   17.8 \\
0750 & 3.0 & 0.071 & 213.0 & 1.355 & &   19.3 \\
0800 & 2.7 & 0.089 & 240.3 & 1.724 & &   39.4 \\
  \hline
\multicolumn{7}{l}{$^a$ Flux density measurement is limited by the ORT confusion.}\\
\multicolumn{7}{l}{$^b$ $\Delta S$ is not limited by confusion and is much accurate.}
\label{tab:tab3}
\end{tabular}
\end{table}

We have used the data obtained from scintillation monitoring observations from 
the ORT for many years towards 0019$-$000 to get the scintillation index as a 
function of heliocentric distance of the point of closest approach of the line 
of sight \citep[see][for details of scintillation index variation with 
heliocentric distance]{mano93,mano06}. During our observation the average 
heliocentric distance was $\sim 185.5{\rm R}_\odot$. Fig. (\ref{f4}) shows the 
long-time average of the scintillation index, and from this, for normal solar 
wind conditions, the expected scintillation index is 0.061 at 
$185.5{\rm R}_\odot$ for a flux density of 3.0 Jy. This corresponds to 
$\langle\Delta S^2\rangle^{0.5}=183$ mJy. Please note that the peak 
scintillation index is $\sim 0.8$ for this source. This implies that the 
source is very compact ($\lesssim 50 $ mas) and is an ideal source for 
scintillation measurements. From our occultation data we can find, using 
Eqn. (\ref{eqn:eqn1}), the normalized level of turbulence 
$(\Phi_{\rm ne}/n_{\rm e}^2)$ in the comet plasma tail compared to that in the 
solar wind. We assume typical values for $L_{\rm s}$ and $L_{\rm c}$ to be 
$10^8$ and $10^5$ km, respectively. Considering that the radio source line of 
sight probes the plasma tail at about $1.4 \times 10^4$ km away from the 
nucleus and that the angular size of the coma was $\sim 3'$, the expected 
value of $L_{\rm c}$ is $\sim 5 \times 10^4$ km. $L_{\rm c}=10^5$ km may be 
taken as an upper limit and, for smaller values of $L_{\rm c}$, the 
corresponding value of $(\Phi_{\rm ne}/n_{\rm e}^2)_{\rm c}$ will increase. 
Average electron number density in the cometary plasma is assumed to be 
$n_{\rm ec}=30$ cm$^{-3}$ \citep{gr86}. Earlier reports \citep{al86,sl86,an87} 
used, following \citet{rm86}, $n_{\rm es}=5$ cm$^{-3}$ for normal solar wind 
condition. We have instead taken the average value of number density to be $7$ 
cm$^{-3}$ computed from the 1 minute OMNI data (from the CDAWeb service of the 
NASA/GSFC Space Physics Data Facility) for the period of 2006 May 21 -- 31. 
We summarize the result of plasma turbulence enhancement in Table 
(\ref{tab:tab3}). 

\section{Discussion}
\label{sec3}

We present the detection of a rare event of the occultation of a compact 
source (B0019$-$000) by the plasma tail of comet Schwassmann-Wachmann 3-B 
using scintillation observation and the power spectral analysis. It is 
interesting to compare the present results with the earlier reports of 
both detections and non-detections of enhanced scintillation. The range of 
values of the normalized plasma turbulence from similar observations spread 
over almost an order of magnitude \citep{al86,sl86,an87,sl90} for a given set 
of parameters (e.g., $L_{\rm c}$, $n_{\rm ec}$). In addition to that, the 
turbulence level is likely to change systematically with the distance from the 
nucleus of the comet along the plasma tail axis. In other words, occultation 
observations probe the plasma at different distances from the comet nucleus and 
hence may reveal a variety of turbulence levels. Hence a straightforward 
quantitative comparison of plasma turbulence is not possible. But from earlier 
results, the clear trend emerges that most of the negative results can be 
attributed to a chance of actually probing a lower level of turbulence in the 
occultation events \citep{sl90} for various reasons. The rest of the cases 
with favorable observing geometry have resulted in a positive detection. In 
the present study, given the possibility of enhanced level of turbulence 
because of the ongoing disintegration and the small offset of the line of 
sight from the nucleus of the comet, our result is in good agreement with this 
scenario. 

The change of the intensity fluctuation power spectra has contributions in two 
distinct scales. The steepening of the spectrum at low frequency indicates 
scale sizes in the intervening plasma at high spatial scales and remarkable 
turbulence enhancement associated in this spectral region. For a typical solar 
wind speed of 400 km s$^{-1}$, it translates to linear scales of $>500$ km. 
At the later part of the observation, at the close encounter of the comet 
nucleus and the line of sight to the radio source, we notice the additional 
significant contribution in the power spectrum at the low-frequency region, 
which corresponds to linear scales of $<50$ km. It is to be noted that the 
IPS measurements normally show much less turbulence at these small scales. The 
evolution of the spectrum with time during the course of the occultation 
indicates that the plasma turbulence level is dominated by small-scale density 
irregularities closer to the tail axis. However, it is the large-scale 
irregularities, developed most likely due to the interaction of plasma with 
the ambient solar wind and the evaporation process, that contribute 
significantly at the outer part of the tail. This shows the presence of two 
main regimes of turbulence in the comet plasma tail. This is found to be in 
agreement with the earlier reported results of \citet{sl86,sl90} for the cases 
of comet Halley and comet Wilson, where similar two-regime plasma turbulences 
were detected.

\section{Conclusions}
\label{sec4}

We have present the results from the occultation observation of a compact 
radio source B0019$-$000 at 326.5 MHz through the plasma tail of comet 
Schwassmann-Wachmann 3-B. We detect significant enhancement in the 
scintillation index and change in the intensity fluctuation power spectrum 
during the occultation period. The systematic study of the evolution of power 
spectrum clearly reveals the dominant effect of large-scale ($>500$ km) 
density fluctuations and the presence of a smaller scale ($<50$ km) closer to 
the tail axis. Since the shape of the power spectra of all the control 
observations show normal quiet solar wind condition, these changes at the 
time of occultation are attributed to the plasma tail of the comet. This clearly 
shows the presence of a two-regime plasma turbulence with two distinct scales 
of density irregularities that may arise from the interaction of solar wind 
with the comet plasma tail. 

\acknowledgments
We thank Jayaram N. Chengalur, C. H. Ishwara-Chandra and Rajaram Nityananda 
for much encouragement and many helpful comments. We are grateful to the 
observing staff of the ORT. One of the authors (P.C.) is thankful to IUCAA for 
sponsoring his research under the IUCAA visiting associateship program. We are 
also grateful to the anonymous referee for useful comments and for prompting 
us into substantially improving this paper. This research has made use of 
NASA's Astrophysics Data System, the JPL Solar System Dynamics Web site, the CDAWeb 
service of the NASA/GSFC Space Physics Data Facility, and the NVSS data products.

\end{document}